# A case study investigation of summer temperature conditions at two coastal sites in the UK, and analysis of future temperatures and heat wave structures in a warming climate scenario.

*Alexandra Edey[1], Ralph Burton[2] and Alan Gadian[2]*

[1] *Office for Nuclear Regulation, Bootle, UK*
[2] *NCAS, University of Leeds, UK*

Correspondence emails:  Alexandra.Edey@onr.gov.uk and Alan.Gadian@ncas.ac.uk

**Abstract:**

Using observational data provided for two UK coastal sites (<3 km from the sea), one on the West coast, in the proximity of the nuclear new build (NNB) site Hinkley Point C (HPC), and the other on the East coast, in the proximity of the proposed NNB site Bradwell B (BRB), changes in surface two-metre temperatures are discussed and compared. The output from the WISER experiment (using the Weather Research and Forecasting (WRF) model) is used for a "control" period, 1990-1995, [*Gadian et al. 2018*]. The nested convective permitting model at a resolution of O(~3km) is driven by a global channel model at a resolution of O(~<20km) which enables a more detailed comparison on the weather scale than is available with current climate models.  Further, using the RCP8.5 warming scenario, the results are compared with the period 2031-2036.  In the future scenario, there is an increase in the number of days where the summer (JJA) model temperatures exceed 25°C. There is a warming of 1.2°C (BRB) and 1.1°C (HPC) in the mean JJA *maximum* daily temperatures for the 2030's model computations compared with the 1990's values, and an average *annual maximum* daily temperature warming of 1.2°C (BRB) and 0.5°C (HPC). For the control period, the model under-predicts both the maximum and the minimum temperatures, particularly the minimum values. Results suggest that there will be a >25% increase in the number of summer days when the maximum temperature exceeds 25°C, a 60% increase when the temperature exceeds the minimum of 13°C in the future scenario, (Tables 3 and 4) and an increase in heat wave temperature events per annum of greater than 10 days, [*Gadian et al. 2018*]. These increases in summer temperatures are larger values than those predicted in the 2013 IPCC assessment [*Collins et al 2013*], but consistent with the maximum temperatures and increased number of hot days listed in the headline figures of UKCP18 [*Met Office 2019(b)*]. These results, with a higher resolution model suggest that the IPCC report is likely to underestimate the increases in maximum temperatures at these two locations.

**1.0 Introduction:**

Temperature is a critical variable for many sectors of industry, e.g. water resources, agriculture, civil protection, energy, transport and tourism, as is precipitation amount, intensity and frequency. Global temperatures are rising [HC826, 2018] and Summer 2019 provided the record highest observed UK temperature (38.7°C) [Met Office, 2019]. The consensus is that extreme high temperatures are becoming more frequent in a warming climate, as suggested in [*Gadian et al.*, 2018 and Met Office, 2019]. Recent papers, [*Berg et al.* 2013 and *Kendon et al.* 2014, 2012] specifically address how "weather" models might respond in a warming climate.

High temperatures are often associated with heat waves. A "heat-wave" is defined as a period of excessively hot weather, which may or may not necessarily be accompanied by high humidity. It is measured in terms of the usual weather in the region and sometimes is relative to the season. Thus temperatures of 25°C may not be considered as very hot in some regions, but in the UK's semi maritime environment it is significant and may cause infrastructure issues. Heat waves can have negative ecological, societal and economic impacts; the 2003 European heat wave affected the majority of French nuclear power plants, in some cases resulting in their temporary shutdown, leading to power interruptions and an increase in electricity prices [*García-Herrera et al.,* 2010]. In summer 2018 several European nuclear power plants, across Finland, Sweden, Germany, France and Switzerland, reduced their output or completely stopped production due to high air and cooling water temperatures [*Nuclear Engineering,* 2018].

In addition to energy, water and food shortages, heat waves can have a deadly impact both directly and indirectly, for example, resulting in forest fires and high pollution levels [*García-Herrera et al.*, 2010]. Studies suggest that heat wave events are likely to increase in frequency over the coming decades [*Russo et al.*, 2015; *Mora et al.*, 2017 and *Gadian et al.*, 2018], with longer dry periods and short (<4 hours) intense precipitation events providing a larger contribution to total precipitation [*Gadian et al.*, 2018].

Blocking anticyclones are typically responsible for heat wave conditions, with Atlantic blocks on average lasting 8-16 days [*Grumm,* 2015]. Persistent anticyclonic conditions, however, produce longer duration heat wave events, such as the 2018 or 2003 European heat wave. The 2003 event saw an unusually warm May-August with temperature anomalies nearly +2°C in May and July, reaching +3.8°C in August and +4.2°C in June (*Black* et al., 2004). Drier than average winter and spring months can also affect summer temperatures, with a severe deficit in soil moisture amplifying temperature anomalies, as the lack of moisture prevents latent heat fluxes from transferring heat upwards [*García-Herrera* et al., 2010; *Black et al.*, 2004]. The 1976 UK heat wave event, which saw 15 consecutive days > 32°C, was preceded by anomalously low precipitation in 1975-1976 (*Russo et al.,* 2015). Sea surface temperature anomalies have also been attributed as a potential factor in amplification of the 2003 European heat wave [*Feudale and Shukla*, 2011]. *García-Herrera et al.* (2010) suggest that the performance of forecasting models could potentially be improved by better characterisation of atmosphere, land surface and sea surface temperature interactions.

In addition to high air temperature, relative humidity and enthalpy should be considered as part of a heat wave event. Both have the ability to affect the Heating, Ventilation and Air-Conditioning (HVAC) plant, alongside temperature. Humidity levels can negatively impact on the cooling capability of a plant, although this is dependent on the cooling mechanism; air conditioning is more challenging in low humidity due to the fact that it cools the air by removing water; by contrast, coolers (cooling humidifiers) are affected by high humidity as they work by adding water to the air. These factors need consideration when trying to mitigate the impacts of summer high temperatures.

Regional convection permitting models are promising tools for improved future climate research [*Prein et al.,* 2015], as they allow convection-scale motions and processes to be simulated [*Nastrom et al.,* 1985]. *Weisman et al.* (1997 and 2000) demonstrate this with the Weather Research and Forecasting Model (WRF) showing that the upper bound on the required resolution is less than 3km. *Gadian et al.* (2018) applied this approach to examine future convective precipitation over the UK and Western Europe. The data produced will be used here. Full details of the approach can be found therein, but a brief description is included below for completeness. In climate models, important meso-scale dynamical structures are sometimes poorly resolved which can lead to limitations for local case study analysis. Long term and quality controlled UK Met Office coastal sites, near the nuclear new build site (NNB), Hinkley Point C (HPC), and the proposed NNB site Bradwell B (BRB) have been chosen for this study. However, the authors believe the results could be representative in the local region. In this paper temperature observations and the model results over the two UK sites for a six year control period (1990-1995) are compared with model results for the same locations (2031-2036). High air and ocean temperatures have recently led to reduced output of European nuclear power plants. This has raised the scientific question, what is the future trend in heat waves likely to be? [*NPR,* 2018 and *Nuclear Engineering,* 2018]

The Intergovernmental Panel on Climate Change (IPCC) report [*Collins et al.,* 2013], Figure FAQ 12.1 (1), suggests that the global temperature rises were ~ < 1°C from 1900 to 1990. The report suggests that there could again be an increase of ~ < 1$^0$C from 1990 to 2030. This paper looks at how the IPCC prediction compares with hourly observations from two UK coastal sites, and the simulations completed using a weather scale convective permitting model.

**2.0 Observational Data:**

The Met Office Integrated Data Archive System (MIDAS) database contains a record of the UK Met Office's existing weather stations, as well the data which they have recorded. The MIDAS data can be accessed via the Centre for Environmental Data Analysis (CEDA) (CEDA, 2018). For this study the 'Met Office – MIDAS Land and Marine Surface Station Data' dataset and the subset 'MIDAS: UK Daily Temperature Data' and 'MIDAS: UK Hourly Weather Observation Data' datasets are used (Met Office, 2006a; 2006b), with the 1990-1995 and 2018 text files downloaded from CEDA. Each text file includes all of the daily temperature or hourly weather observation data recorded at all operating weather stations in the UK for that year.

The CEDA online search for Met Office MIDAS stations (CEDA, 2018) is used to find the weather stations which recorded data for the periods of interest (1990-1995 and 2018) and which were in the proximity of the HPC or BRB sites, Figure 1(a). The selected MIDAS stations were:

- Minehead (src_id 1288) – this extant weather station operated between 01/01/87-31/12/95 and was located approximately 24km from the HPC site. The data from this weather station was used to look at temperature observations 1990-1995 near HPC.

- Shoeburyness: Landwick (src_id 498) – this weather station has operated since 01/01/81 and is located approximately 24km from the BRB site. The data from this weather station was used to look at temperature observations 1990-1995 near BRB.

- St Athan (src_id 19206) – this weather station has operated since 01/09/1988 and is located approximately 29km from the HPC site. The data from this weather station was used to look in depth at a heat wave event in 2018.

The choice of weather station was limited by geographical location and the date of operation. Unfortunately, the Minehead station did not record data from the beginning of May 1995 to end of December 1995 and so could not be used for comparison during this period.

### 3.0 Model setup:

A nested regional model structure is used for these simulations.   Full details of the Weather Impact Study at Extreme Resolution (WISER) dataset and the Weather Research Forecasting (WRF) model (3.5.1), based on the RCP 8.5 forcing scenario, that was used for this study can be provided in *Gadian et al*. (2018), but a summary is discussed below.  High resolution global models are too resource expensive to emulate "weather" on climate timescales, and the regional modelling provides a compromise.  *Done et al.* [2015] uses the WRF model to demonstrate the value of the nested approach in capturing meso-scales for the case of tropical cyclones in the West Atlantic. Figure 1(b) displays the domain structure used, with resolution from < 3.2km and 2.0km at 35°N and 68°N in the inner domain (d02) increasing by a factor of five for the outer domain, implying resolutions ~O(3km) and ~O(12km). 51 vertical stretched levels are used, with a lid at 10 Hectopascals. 1731 East - West grid points in both domains and 907 and 1001 grid points in the North - South directions in the outer and inner domains, respectively, provide the required discretisation. Convective parameterisation is not applied in the inner domain, but the Kain-Fritsch (K-F) [*Done et al.* 2015, *Klemp et al.* 2006, *Weisman et al.* 2008] scheme is used in the outer domain. The Yonsei University (YSU) boundary layer scheme, WRF-single-moment-microphysics class 6 (WSM6) and a 5 layer Noah land surface schemes [*Weisman et al.* 2004, 2008] are used.

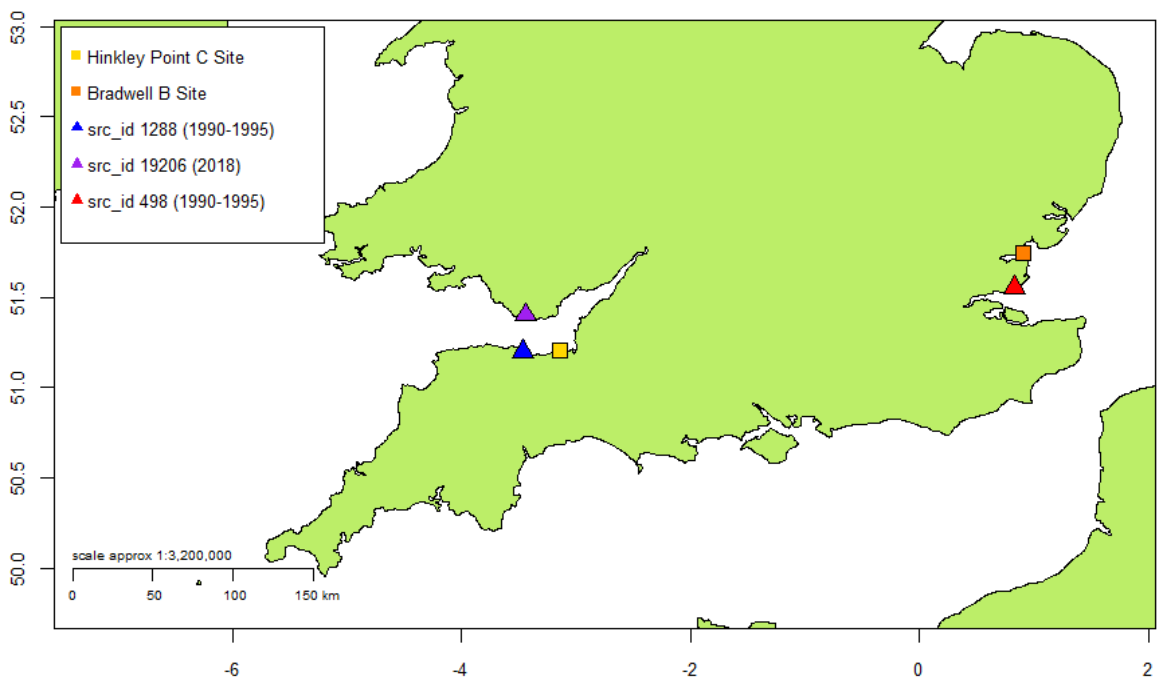

*Figure 1a.*

Map of the location of observational sites (triangles) used for data comparison with the location of the model site positions (squares)

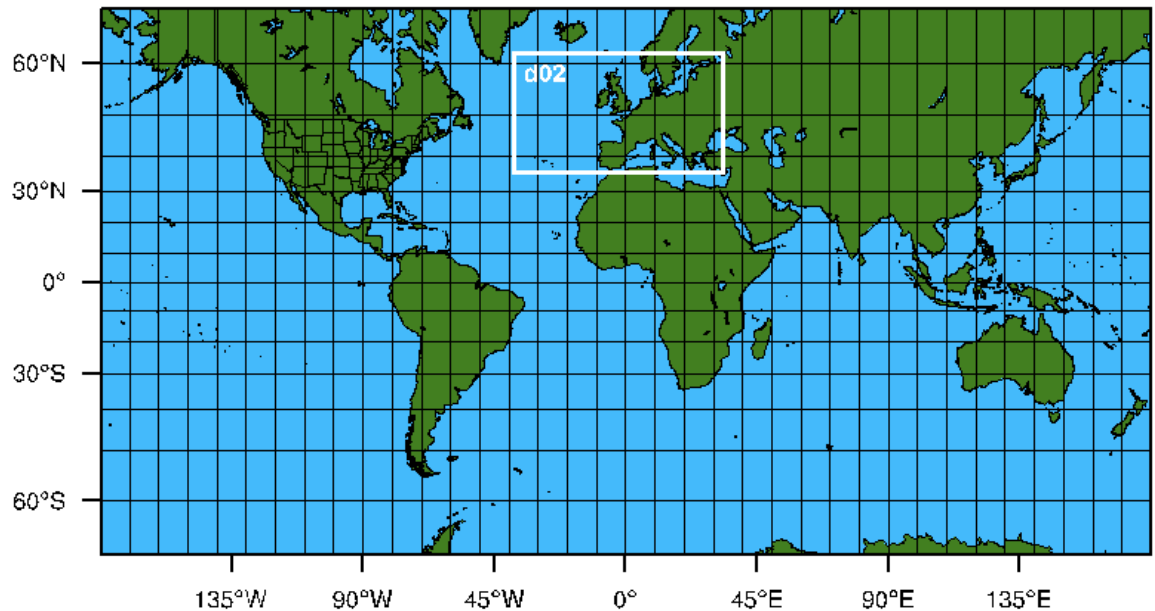

*Figure 1b.*

Domain structure for the simulation. The outer domain (d01) resolution is 20km at +/- 30° and 8km at 68° N/S. The inner domain (do2) is one way nested at a ratio of 5:1 (taken from Gadian et al. 2018).

Surface and lateral boundary forcings drive the model outer domain (d01) for the lower boundary sea surface (SST) condition and the northern and southern channel boundaries. The two domains are one way nested, with the boundaries of domain d02 driven by d01 data. Although ERA-interim data were used for the control runs (1989–1995), only Community Earth System Model (CESM) version 1 data are utilised for both the control run (1989–1995) and for the future run (2030–2036). There was only a small difference between the two sets of results and so for brevity only the results CESM driving boundary data are discussed. The boundary data are bias corrected before being used to drive WRF; the bias correction approach is justified [*Bruyere et al*, 2014] with a detailed technical description [*Bruyere et al*, 2015] of the CESM data [*Monaghan et al*, 2014] all detailing the manipulation of the required input parameters for the boundary conditions. The approach corrects for the mean error in the simulated climate by replacing CESM climatology with reanalysis climatology, but retains the daily weather (6 hourly) and longer period weather variability from CESM. Surface hourly data, including 2 metre relative humidity, temperatures, 10 m wind speed and direction, rainfall, skin temperature are stored at 1 hourly intervals. Data produced for the first year of each simulation, (1989 and 2030), are ignored to remove any model spin up effects.

For this study, the coordinates of the HPC and BRB sites were used to extract the relevant WISER data and to allow comparison with the MIDAS data detailed in Section 2.

### 4.0 Results and Discussion:

The purpose of this study is to examine the predicted future changes in summer temperatures and the relative ability of the model to represent these temperatures. The first section will examine the relative changes by comparing 1990-1995 observational data with the 1990-1995 control (model) data.

**4.1 Annual and Summer Maximum and Minimum Daily Temperatures:**

*Comparison of Summer 1990-1995 Observational and Model Data*

The summer (June, July and August) hourly MIDAS temperatures are compared with those produced in the model for the control period, 1990-1995, with box plots for BRB (Figure 2(a)), for HPC (Figure 2(b)) and the average maximum daily temperature (Table 1). The MIDAS dataset for the HPC location omitted JJA values for 1995 and so only 1990-1994 is used for Figure 4.2. The grid points used were those closest to the meteorological stations.

All Numerical Weather Prediction models contain biases due to their coarse resolution and limitations in the representation of physical processes. As these biases can adversely affect the dynamical downscaling results, it is important to first bias-correct model data. There are several approaches that can be used, but the approach of (Bruyère et al., 2014 and 2015) is typical. This method uses global and surface reanalysis data to correct the mean bias in the global fields. The bias correction method only corrects the mean state, whilst the synoptic and climate-scale variability is maintained. For future, and current weather models (e.g. *Gadian et al.,* 2018) this can be achieved, for example, by combining a 20-year mean annual cycle from the reanalysis data with the 6-hourly perturbation terms from the model being run. For example, commonly, when applying models to look at specific site variables, such as temperature, the mean offset of the model compared to observations needs to be considered, before applying predictions for future values.

For the control run, the output for both sites suggests that the model underestimates the mean summer maximum daily observational temperatures by 1.8°C and 0.5°C, and the night time (minimum) temperatures by 1.8°C and 2.8°C, for the BRB and HPC localities respectively. This provides an offset which can be used to apply to the future model temperatures. BRB is on the east coast, where it could be argued the maritime Sea Surface Temperatures have less influence, with BRB exhibiting a slightly more continental climate (Table 1(a)).

*4.2 Comparison of Summer 1990-1995 and 2031-2036 Model Data*

For the future scenario, the model mean summer *maximum* daily temperature increases by 1.2°C and 1.1°C for BRB and HPC, with little change in the standard deviation (Table 1(a)) and the minimums for summer *maximum* daily temperature increase by 2.0°C and 3.0°C. The model mean summer *minimum* daily temperature increases by 1.3°C and 1.1°C for BRB and HPC (Table 1(b)). Interestingly, the summer *maximum* daily temperature quartile 1 range increase of 1.6°C and 1.2°C for BRB and HPC is larger than the increase in the mean summer *maximum* temperatures (Table 1(a)). These results are consistent with *Gadian's et al.* (2018) conclusion that there are longer dry spells, shorter heavier precipitation events and with an increased frequency of warm temperatures. The weather model used in this study under-estimates the warming locally, as the WISER mean temperatures always appear less than the observations at these sites. There is no reason to suggest that this result will not be duplicated in the future scenarios. Consideration should be given as to whether other lower resolution models may similarly under-estimate local warming.

*4.3 Comparison of Annual Data*

The annual average maximum and minimum changes at the two sites show a similar signal (Table 2). The underestimate of the WISER model annual *maximum* temperatures compared with the MIDAS observational data for the control period (1990-95) is 1.5°C and 0.6°C for BRB and HPC, with corresponding *minimum* temperature underestimates of 1.2°C and 2.1°C.

For a comparison of the annual average changes in the control and the future model simulations, there is a warming of 1.2°C and 0.5°C for BRB and HPC, with the minimums increasing by 0.1°C and 0.6°C. These results suggest that the summer mean extreme temperatures show a bigger signal of warming, but given the standard deviation of ~ 3.2°C it cannot be taken as significant with this limited data set.

*5 Discussion*

The authors are assuming, as a first approximation, that the underprediction of extreme temperatures for the control (1990-1995) WISER model simulations will apply also to the 2031-2036 WISER results. Using linear extrapolation on the WISER data, for HPC, one could anticipate that the increase in mean summer *maximum* daily temperature to be ~ 2.2$^0$C warmer in 2070s c.f. values for 1990s (~1.1°C warmer in 2030's cf. 1990s). Linear extrapolation is a conservative assumption, i.e. this, in the opinion of the authors, may be an underestimation of what could be expected as the rate of warming is likely to increase over time. Additionally, for BRB, one could expect the increase in mean summer *maximum* daily temperature to be ~ 2.4°C warmer in 2070s c.f. values for 1990s (~ 1.2°C warmer in 2030's c.f. 1990's). Potentially, based on data from this study, the sites would be at least ~3°C warmer by 2100's than the 1990's and ~ 4°C warmer than preindustrial times; this is produced using a conservative linear extrapolation.

Amongst other factors, this assumes that there is continuance of the RCP8.5 radiational driving effect, and that there is no change to the Atlantic Meridional Ocean Circulation (AMOC) in the North Atlantic, which is considered unlikely by Ceasar et al. 2018. However, *Collins et al.* (2013) indicates the same warming trend until 2030 for even the RCP2.6 case (the lowest case scenario with a radiational forcing increase of only 2.6 W/m$^2$). Based on the model used in this study, the authors consider that at the two sites a 2°C warming above pre-industrial is likely to be an underestimate. Our calculations suggest that it is feasible that the summer temperature warming could be ~1°C between the control period (1990-1995) and the future (2031-2036) period. Specifically for HPC (*BRB*) the difference in summer *maximum* daily temperatures is 1.1°C (*1.2°C*) and for summer *minimum* daily temperatures is 1.2°C (*1.3°C*).

If these model results are realistic representations, then a greater summer maximum temperature warming is suggested at these sites in the UK than is produced by the IPCC report [*Collins et al.* 2013]. The report, section 11.3, states "*The global mean surface temperature change for the period 2016–2035 relative to 1986–2005 will likely be in the range of 0.3°C to 0.7°C (medium confidence)*". These figures are smaller than the WISER model data presented in the previous two paragraphs. The WISER average mean *maximum* daily temperatures are larger. As explanation, it should be appreciated that the IPCC climate models generally have resolutions larger than 60km and parameterise convection. The UK Climate Projections 2018 (UKCP18) observations at ~ 2.24km state that "*The average temperature over the most recent decade (2009-2018) has been on average 0.3°C warmer than the 1981-2010 average and 0.9°C warmer than the 1961-1990 average*", (UK Met office 2019 and 2019b). For the UK, the 0.3°C figure is consistent with WISER mean predictions for the 2020-2025 simulation, not yet published, when compared with 1989-1995 average, these results are not presented for brevity. The WISER model is a single study of resolution ~ 2-3kmthat permits convection and therefore could be expected to exhibit different characteristics, especially for extreme temperatures and local scale features and, by comparison, the IPCC report is likely to underestimate the increases in maximum temperatures at these two locations.

However, the magnitude of these increases is similar in the UKCP18 2.2km ensemble calculations [*Kendon et al., 2019* and *UK MET Office, 2019b*]. For the UK average, hot summer days will warm more, with a range of 3.7 to 6.8°C, and summer averages are predicted to be 3.6-5.0°C hotter by (2061-2080). This is consistent with the conservative linear extrapolation estimates of the WISER data.

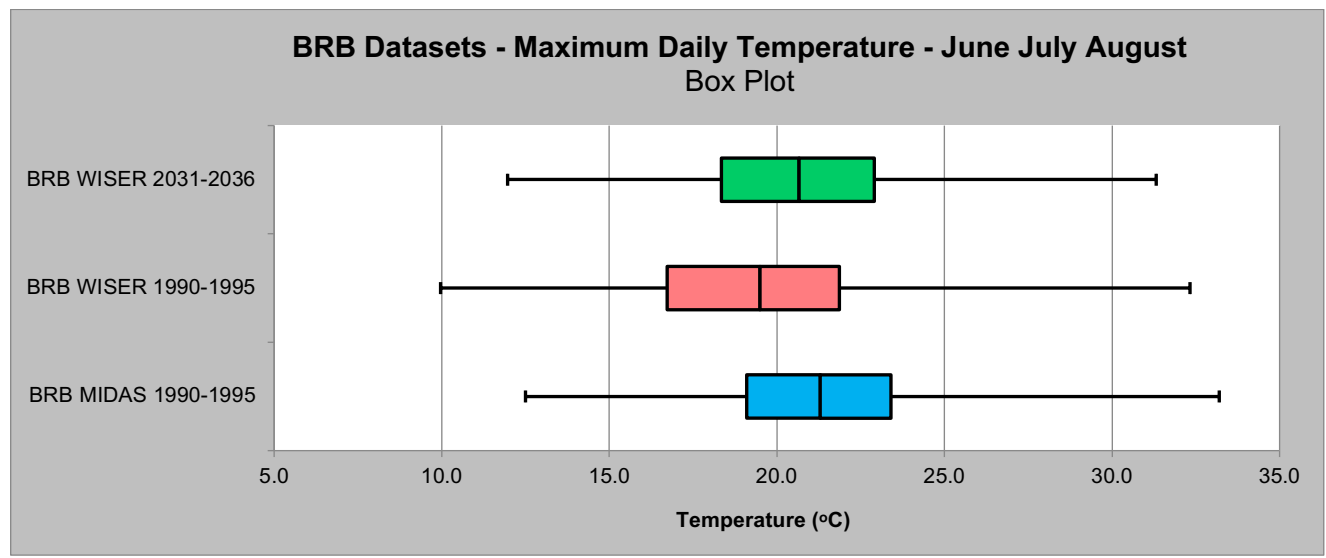

**Figure 2(a)** Plot showing the: minimum, quartile 1, median, quartile 3 and maximum values for the maximum daily temperatures in June July and August for each BRB data set, with no bias removal.

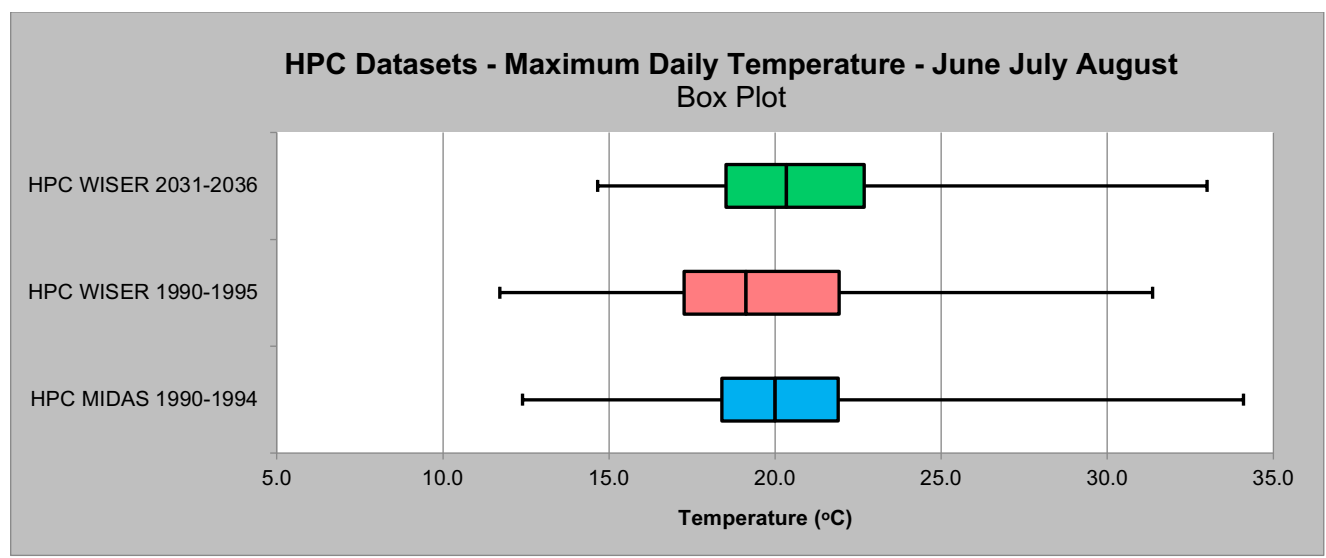

**Figure 2(b)** Plot showing the: minimum, quartile 1, median, quartile 3 and maximum values for the maximum daily temperatures in June July and August for each HPC data set.

|  | HPC | | | BRB | | |
|---|---|---|---|---|---|---|
|  | MIDAS 1990-1994 | WISER 1990-1995 | WISER 2031-2036 | MIDAS 1990-1995 | WISER 1990-1995 | WISER 2031-2036 |
| Mean | 20.2 | 19.7 | 20.8 | 21.2 | 19.4 | 20.6 |
| SD | 2.8 | 3.5 | 3.2 | 3.3 | 3.7 | 3.3 |
| Minimum | 12.4 | 11.7 | 14.7 | 12.5 | 10.0 | 12.0 |
| Q1 | 18.4 | 17.3 | 18.5 | 19.1 | 16.7 | 18.3 |
| Median | 20.0 | 19.1 | 20.3 | 21.3 | 19.5 | 20.7 |
| Q3 | 21.9 | 21.9 | 22.7 | 23.4 | 21.9 | 22.9 |
| Maximum | 34.1 | 31.4 | 33.0 | 33.2 | 32.3 | 31.3 |

**Table 1(a)** Data for the average June July August Maximum Daily Temperature Box Plots.

|  | HPC | | | BRB | | |
|---|---|---|---|---|---|---|
|  | MIDAS 1990-1994 | WISER 1990-1995 | WISER 2031-2036 | MIDAS 1990-1995 | WISER 1990-1995 | WISER 2031-2036 |
| Mean | 12.6 | 9.8 | 10.9 | 12.5 | 10.7 | 12.0 |
| SD | 2.3 | 2.4 | 2.2 | 3.1 | 2.4 | 2.3 |
| Minimum | 2.7 | 1.7 | 3.7 | 2.5 | 3.4 | 5.1 |
| Q1 | 10.9 | 8.4 | 9.5 | 10.4 | 9.1 | 10.6 |
| Median | 12.8 | 9.7 | 11.1 | 12.9 | 10.8 | 12.0 |
| Q3 | 14.2 | 11.4 | 12.5 | 14.8 | 12.2 | 13.6 |
| Maximum | 18.3 | 16.5 | 16.7 | 20.4 | 16.9 | 18.0 |

**Table 1(b)** Data for the average June July August Minimum Daily Temperature Box Plots.

| Location | Data Type | Average Daily Maximum Temp. °C | Average Daily Minimum Temp. °C |
|---|---|---|---|
| HPC | MIDAS 1990-1995 | 14.2 | 7.8 |
|  | WISER 1990-1995 | 13.6 | 5.7 |
|  | WISER 2031-2036 | 14.1 | 6.3 |
| BRB | MIDAS 1990-1995 | 14.4 | 7.4 |
|  | WISER 1990-1995 | 12.9 | 6.2 |
|  | WISER 2031-2036 | 14.1 | 6.3 |

**4.2 Summer Hourly Temperature, Wind Speed and Humidity:**

Examination of the hourly summer (June, July, August) temperatures is important to establish the daily heating cycle on a building structure. Unfortunately, hourly temperatures are not available for the control period at either site, but recent hourly values in the vicinity of the HPC site are available and these are compared with the model hourly variables for a summer during the control period and one in future time series. Humidity and wind speed values are also provided.

In Figure 3(a), the 2018 JJA hourly measurements are plotted for St Athan, across the Bristol Channel from the HPC site. A heat wave can be observed between the 25th June and 4th July. The temperature trace (red) indicates maximum values above ~ 25°C for this period. This period will be examined in more detail in the next section. The humidity trace (green) shows a corresponding reduction in relative humidity to ~ 40% at the hottest part if the day and remaining below 80% at the coolest part of the night for this period, with an average value of ~ 55%. Normally the humidity averages about 70%. For much of the summer, the wind speeds vary between 2 and 10m/s, wind speeds are often lower during heat wave events but this is not observed for the end June-early July heat wave.

Figures 3(b) and 3(c) show the same plots for summer 1995 and 2035, using the WISER data. Both years were selected as they were relatively warmer years and so are most likely to be considered 'heat wave years'. This means that 1995 and 2035 cannot be considered representative years for the 1990's or 2030's but they do allow for comparison of heat wave events. In the 1995 data, Figure 3(b), there is a heat wave structure at the beginning of July, with characteristic humidity, temperature, structure and wind speed. Comparison with the 2035 data, Figure 3(c), suggests that the structure of the summer heat waves remains the same within the model data. Comparing Figures 3(b) and 3(c) there is suggested further, albeit limited, evidence of temperature increases in future decades. As shown in *Gadian et al.* (2018), the data here is consistent with the result that there are longer periods of low wind speeds.

The UKCP18 report addresses future changes in hourly temperatures and will be available for analysis [*UK Met Office,* 2019b]. Similarly they find a future summer decrease in soil moisture, consistent with reduced rainfall and consider that *"locally this could lead to an exacerbation of the severity of hot spells",* with increases in hourly convective precipitation rates. Hot spells, defined as two days at more than 30°C, rise from an average of 0.25 occurrences per year in the present-day to 4.3 by 2070 in UKCP Local (2.2km), as discussed further in section 4.3.

When comparing the 2018 MIDAS data (Figure 3(a)) with the WISER data (Figures 3(b) and 3(c), as discussed above, the WISER data under predicts summer night time temperatures. The WISER data difference between day time and night time temperatures is notably larger than the 2018 MIDAS data. This is again consistent with section 4.1. Care needs to be taken when comparing the datasets because the WISER data does not cover the 2018 heat wave. However, this is not included here because it merits further investigation and study.

The observational data and model results show consistent increases in wind speed at the hottest part of the day, although the timing within an hour or so, does not necessarily match, presumably because representation of boundary layer dynamics in numerical models is still subject to much development.

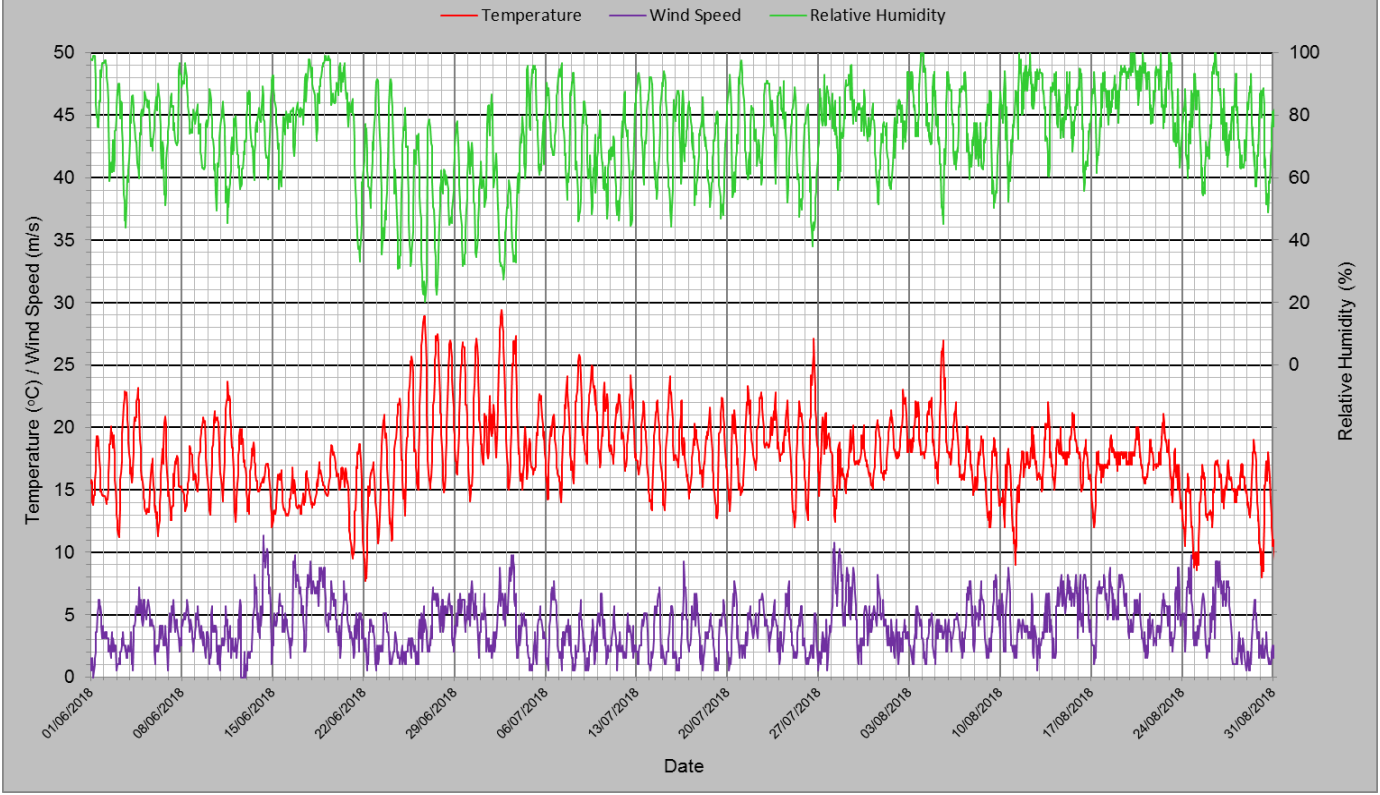

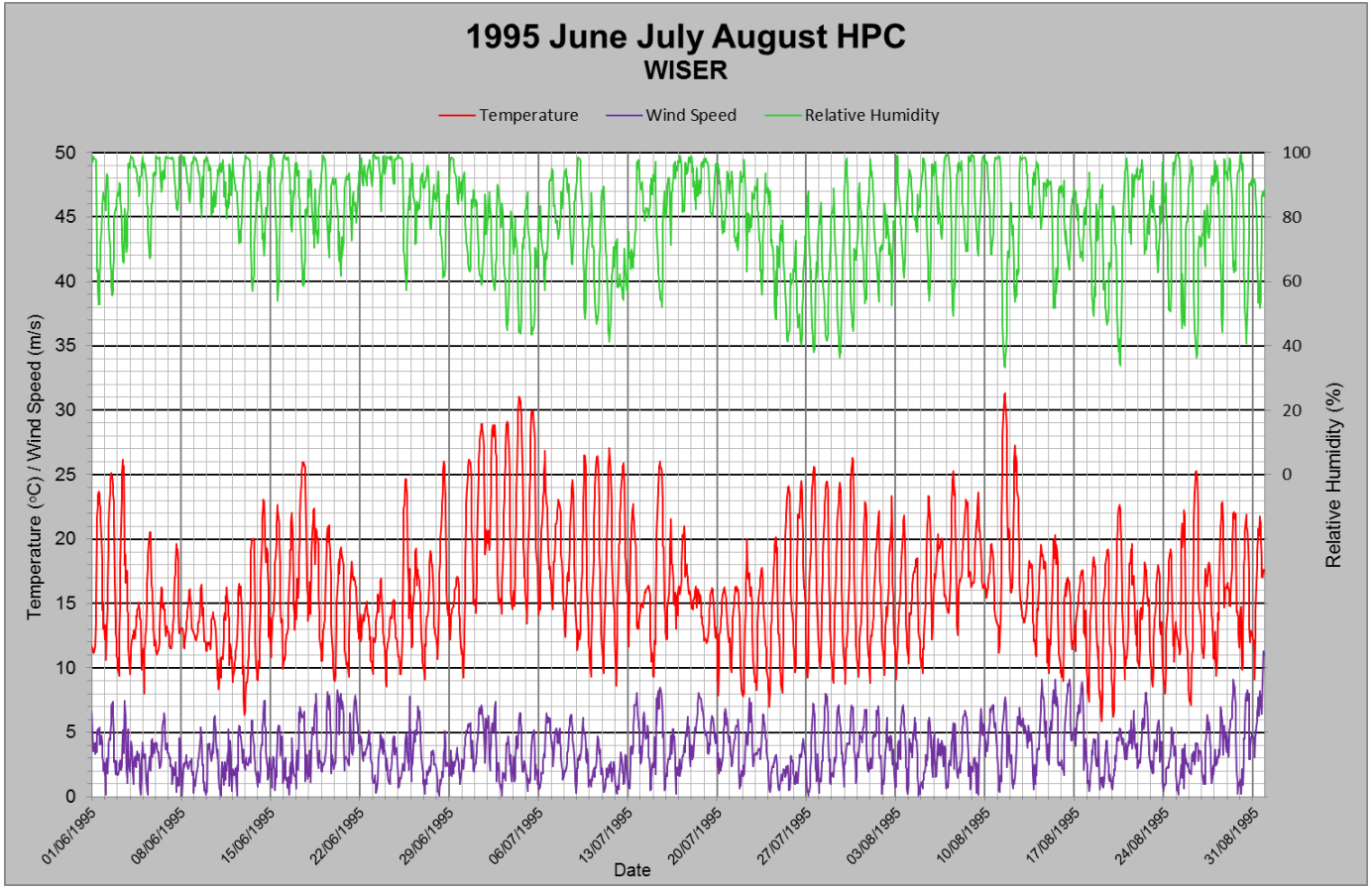

**Figure 3(b)** Temperature, wind speed and humidity data for June, July and August WISER 1995.

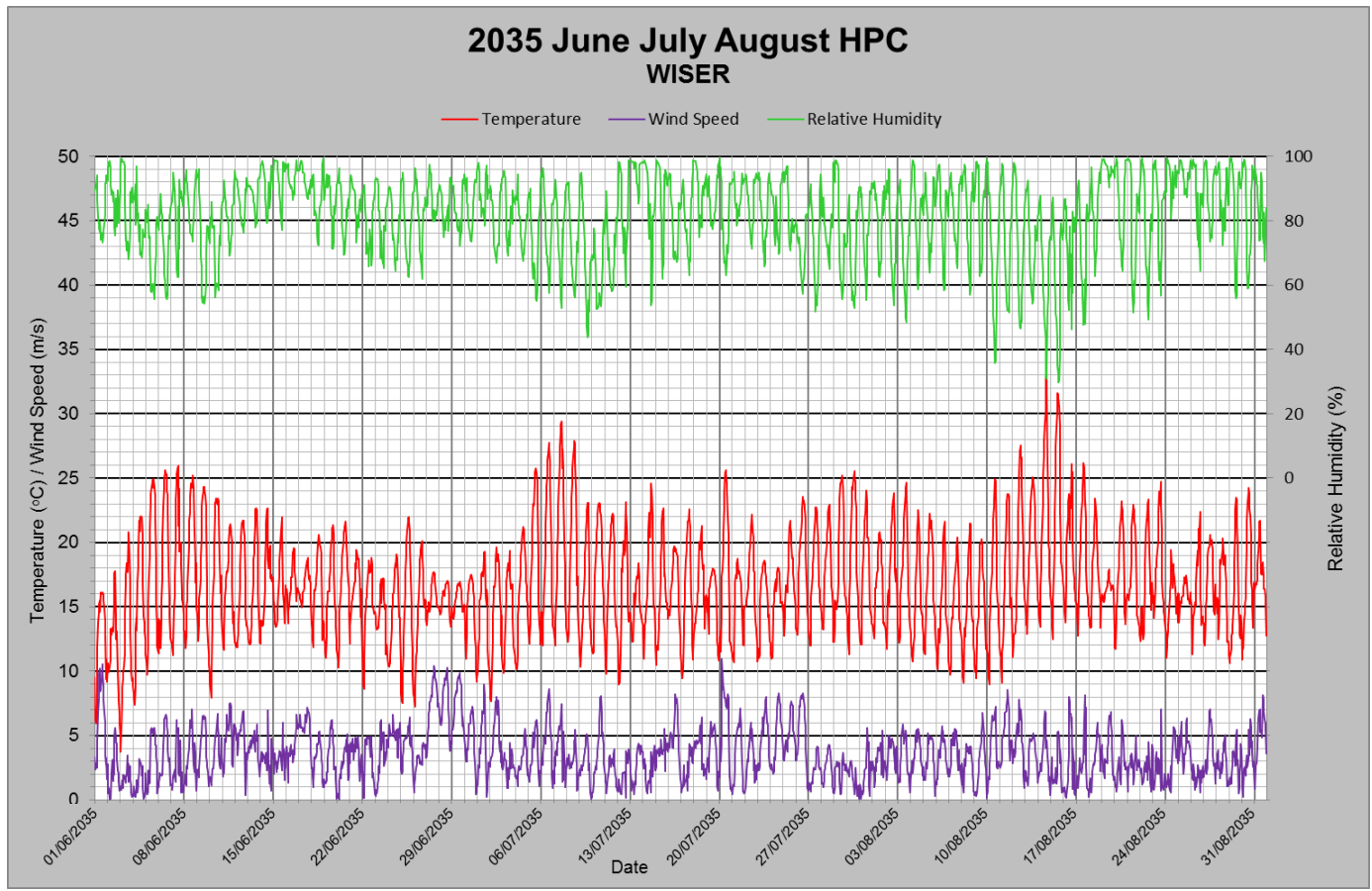

**Figure 3 (c)** Temperature, wind speed and humidity data for June, July and August WISER 2035.

**4.3 Heat Wave Events and Structures:**

A "heat-wave" is defined as a period of excessively hot weather (of unspecified length), which may or may not necessarily be accompanied by high humidity. It is measured in terms of the usual weather in the region and sometimes is relative to the season. Thus temperatures of 25°C may not be considered as very hot in some regions, but in the UK's semi maritime environment it is significant and may cause infrastructure issues. The UK Met Office operates a Heat Health Watch system which differs for each Local Authority region (*Heat wave definition*, 2019). Heat wave conditions are defined by the maximum daytime temperature and minimum night-time temperature rising above the threshold for a particular region; the minimum is taken as 15°C and the maximum 30°C / 25°C. For the sake of this paper, however, we have defined a heat wave occurring when daytime temperatures exceed 25°C and a "nominal minimum" of 15°C. As discussed earlier in section 4.1, the WRF model appears to underestimate the summer night-time minimum by up to 2.8°C, so the definition applied in this study is when the night-time "actual minimum" >13°C (as opposed to 15°C) and a daytime maximum >25°C. An undocumented list of UK heat wave conditions has been suggested by the authors (e.g. 1955, 1976, 1990, 1995, 1997, 2003, 2007, 2011, 2015, 2017, 2018 and 2019) but the authors could not find a list available in the peer reviewed literature. However, the results presented argue that these WISER weather model calculations show a future significant increase in heat wave conditions, larger than those predicted in the 2013 IPCC assessment [*Collins et al.,* 2013].

This section will discuss the frequency of heat wave events in the observations and in the WISER model results.

Unfortunately there was no hourly observational data available for the control period (1990-1995). However, Figure 4 focuses on a heat wave event in 2018 recorded by the St Athan weather station. The structure of the heat wave event is evident, the temperature builds over a few (3-4) days before peaking. A day time temperature >25°C is maintained for 6 days before it dips for a single day and then returns to >25°C daytime temperature for two further days before temperatures once again drop. Day time maximum wind speeds generally peak at ~ 9 m/s, normally < 6m/s reducing to 3m/s at night and the humidity inversely mirrors the temperature. In the UK, the humidity values may reach 80% in coastal regions, due to the evaporation of sea water, which often has a cooling effect. This is discussed in *Harvis (2005)*, and with particular reference to the US, where coastal currents can be influential.

Heat wave patterns are replicated in the WISER model results, as seen in Figures 3(b) and 3(c) (at the beginning of July 1995 and the beginning of July and mid-August in 2035). A corresponding detailed plot of the WISER data, equivalent to Figure 4, has been omitted for brevity reasons. It provides no more insight and the patterns are similar in the observations and the WISER model values.

The heat wave plot, Figure 4, also has a close resemblance to the 2003 UK heat wave event, taken from *Black et al.* (2004). Both show similar temperature fluctuations from day to night time and show day time temperatures repeatedly exceeding >25°C. The paper also displays the increase in ground heat flux, with the dry ground surface strongly emitting sensible heat from a site at Reading University. This is a factor in maintaining the heat wave.

The signal regarding the heat wave frequency, as in Table 3 (HPC) is only partly clear, mainly due to some missing observational data. For the whole control period (1990-1995) (Table 3a), the heat waves of 1990 dominate the frequency data set. 1995 was also a hot summer, but unfortunately, no data is available for HPC. The effect of 1995 is apparent in Table 4 (BRB), where the consequences of the heat wave frequency are more apparent.

In Table 3a, excluding 1995 due to lack of observations, the observations produce 20 days of temperatures over 25°C, and the WISER model (Table 3b) provides 27 days. The numbers of days which satisfy the heat wave condition (maximum daily temperature >25°C and minimum daily temperature >13°C) amount to 18 days for the observations, and 4 days for the WISER model data, largely because of the low minimum temperatures at the time of high maximums. Although this is a very short period for making a comparison, and therefore quite speculative, for the 5 years (excluding summer 1995), there were two events when the heat wave condition was satisfied in both the observations and the model (Tables 3(a) & (b)). The underestimation of the WISER night-time minimum temperatures remains an issue with this comparison and the accommodation of the negative offset remains a challenge for this analysis.

When comparing the 6 year 1990-1995 WISER data set with the values for 2031-2036, (Tables 3b and 3c) there is a clear signal. Even though 1995 was a hot year, there is a noticeable increase in heat wave conditions in the future scenario range. The number of days when the temperature exceeds 25°C increases from 48 to 63, the number of days when the minimum temperature exceeds 13°C increases from 51 to 95 and the number of 4 day periods when the heat wave condition applies, increases from 4 to 7.

In Table 4, the summer data for BRB is displayed. The advantage of this example is that there is a full set of 6 year met. station observations at Shoeburyness. There were 67 days when temperatures exceeded 25°C, 274 days when the minimum exceeded 13°C, 61 days when heat wave conditions apply and 3 for consecutive 4 day events, in years 1990 and 1995 (Table 4a). The corresponding values for the WISER results are 33 days where the maximum temperature > 25°C, 86 days when the minimum temperature > 13°C, 15 days when heat wave conditions and no 4 day events (Table 4b). This is consistent with the conclusions in section 4.1 that the WISER model underestimates the higher temperatures.

However, comparing the Table 4b and 4c for the control and future climate, the number of days when the maximum temperature exceeds 25°C increases from 33 to 49, the number of days when the minimum exceeds 13°C increases from 86 to 189, and the number of heat wave days increases from 15 to 27. Applying the offset, described in the paragraph above, that the WISER model underestimates the number of hot days and warm nights by a factor of two or three, to the 2031-36 model results produces a relatively large change. A linear and conservative projection of the control results to the future 2031-2036 could also give an increase of more than 25% of days each year when the maximum temperature > 25°C, over 60% increase of events when the minimum temperature is exceeded, with at least 10 days every year when heat wave type temperatures would exist at this location. *Gadian et al.* (2018) also suggested that there would be an increase of 2 hours a day in the occurrence of wind speeds less than 3 m/s.

**Table 3(a-c)**

**MIDAS - Minehead (src_id 1288) - June/July/August**

| | Year | Number of days | | | Number of times |
|---|---|---|---|---|---|
| | | Max temp >25 | Min temp >13 | Max temp > 25 + Min temp >13 | ≥ 4 consecutive days max temp >25 |
| 1 | 1990 | 15 | 46 | 12 | 2 |
| 2 | 1991 | 2 | 37 | 2 | 0 |
| 3 | 1992 | 2 | 47 | 3 | 0 |
| 4 | 1993 | 0 | 34 | 0 | 0 |
| 5 | 1994 | 1 | 53 | 1 | 0 |
| 6 | 1995* | 0 | 0 | 0 | 0 |
| | Total | 20 | 217 | 18 | 2 |
| | Average | 3.3 | 36.2 | 3.0 | 0.3 |
| | Total (excl. yr 6) | 20 | 217 | 18 | 2 |
| | Average (excl. yr 6) | 4.0 | 43.4 | 3.6 | 0.4 |

*data from Jan-May ONLY*

**Table 3(a)** Table of the number of JJA days when the maximum temperature exceeds 25°C and the minimum exceeds 13°C. (a) Refers to observational data from Minehead.

**WISER - HPC - June/July/August**

| | Year | Number of days | | | Number of times |
|---|---|---|---|---|---|
| | | Max temp >25 | Min temp >13 | Max temp > 25 + Min temp >13 | ≥ 4 consecutive days max temp >25 |
| 1 | 1990 | 4 | 9 | 0 | 0 |
| 2 | 1991 | 8 | 15 | 4 | 1 |
| 3 | 1992 | 3 | 4 | 0 | 0 |
| 4 | 1993 | 11 | 7 | 0 | 1 |
| 5 | 1994 | 1 | 1 | 0 | 0 |
| 6 | 1995 | 21 | 15 | 8 | 2 |
| | Total | 48 | 51 | 12 | 4 |
| | Average | 8.0 | 8.5 | 2.0 | 0.7 |
| | Total (excl. yr 6)† | 27 | 36 | 4 | 2 |
| | Average (excl. yr 6)† | 5.4 | 7.2 | 0.8 | 0.4 |

†*for direct comparison with 3(a) data*

**Table 3(b)** Table of the number of JJA days when the maximum temperature exceeds 25°C and the minimum exceeds 13°C. (b) Refers to the HPC WISER data for 1990-1995.

**WISER - HPC - June/July/August**

| | Year | Number of days | | | Number of times |
|---|---|---|---|---|---|
| | | Max temp >25 | Min temp >13 | Max temp > 25 + Min temp >13 | ≥ 4 consecutive days max temp >25 |
| 1 | 2031 | 17 | 25 | 5 | 2 |
| 2 | 2032 | 11 | 18 | 3 | 1 |
| 3 | 2033 | 7 | 14 | 3 | 0 |
| 4 | 2034 | 8 | 7 | 0 | 0 |
| 5 | 2035 | 16 | 15 | 2 | 2 |
| 6 | 2036 | 4 | 16 | 2 | 2 |
| | Total | 63 | 95 | 15 | 7 |
| | Average | 10.5 | 15.8 | 2.5 | 1.2 |
| | Total (excl. yr 6)[†] | 59 | 79 | 13 | 5 |
| | Average (excl. yr 6)[†] | 11.8 | 15.8 | 2.6 | 1 |

[†]*for direct comparison with 3(a) data*

**Table 3(c)** Table of the number of JJA days when the maximum temperature exceeds 25°C and the minimum exceeds 13°C. (c) Refers to the HPC WISER data for 2031-2036.

**Table 4 (a-c)**

**MIDAS - Shoeburyness: Landwick (src_id 498) - June/July/August**

| | Year | Number of days | | | Number of times |
|---|---|---|---|---|---|
| | | Max temp >25 | Min temp >13 | Max temp > 25 + Min temp >13 | ≥ 4 consecutive days max temp >25 |
| 1 | 1990 | 15 | 41 | 14 | 1 |
| 2 | 1991 | 6 | 49 | 6 | 0 |
| 3 | 1992 | 6 | 50 | 5 | 0 |
| 4 | 1993 | 6 | 30 | 6 | 0 |
| 5 | 1994 | 9 | 55 | 8 | 0 |
| 6 | 1995 | 25 | 49 | 22 | 2 |
| | Total | 67 | 274 | 61 | 3 |
| | Average | 11.2 | 45.7 | 10.2 | 0.5 |
| | Total (excl. yr 6)[†] | 42 | 225 | 39 | 1 |
| | Average (excl. yr 6)[†] | 13.4 | 45.0 | 7.8 | 0.2 |

[†]*for direct comparison with 3(a) data*

**Table 4a** Table of the number of JJA days when the maximum temperature exceeds 25°C and the minimum exceeds 13°C. (a) Refers to observational data from Shoeburyness.

**WISER - BRB - June/July/August**

| | Year | Number of days | | | Number of times |
|---|---|---|---|---|---|
| | | Max temp >25 | Min temp >13 | Max temp > 25 + Min temp >13 | ≥ 4 consecutive days max temp >25 |
| 1 | 1990 | 2 | 13 | 1 | 0 |
| 2 | 1991 | 7 | 18 | 4 | 0 |
| 3 | 1992 | 4 | 4 | 1 | 0 |
| 4 | 1993 | 9 | 9 | 2 | 0 |
| 5 | 1994 | 0 | 14 | 0 | 0 |
| 6 | 1995 | 11 | 28 | 7 | 0 |
| | Total | 33 | 86 | 15 | 0 |
| | Average | 5.5 | 14.3 | 2.5 | 0.0 |
| | Total (excl. yr 6)[†] | 22 | 58 | 8 | 0 |
| | Average (excl. yr 6)[†] | 4.4 | 11.6 | 1.6 | 0 |

[†]*for direct comparison with 3(a) data*

**Table 4(b)** Table of the number of JJA days when the maximum temperature exceeds 25°C and the minimum exceeds 13°C. (b) Refers to the BRB WISER data for 1990-1995.

**WISER - BRB - June/July/August**

| Year | | Number of days | | | Number of times |
|---|---|---|---|---|---|
| | | Max temp >25 | Min temp >13 | Max temp > 25 + Min temp >13 | ≥ 4 consecutive days max temp >25 |
| 1 | 2031 | 12 | 44 | 8 | 0 |
| 2 | 2032 | 1 | 22 | 1 | 0 |
| 3 | 2033 | 2 | 29 | 0 | 0 |
| 4 | 2034 | 12 | 27 | 6 | 0 |
| 5 | 2035 | 13 | 39 | 8 | 0 |
| 6 | 2036 | 9 | 28 | 4 | 0 |
| Total | | 49 | 189 | 27 | 0 |
| Average | | 8.2 | 31.5 | 4.5 | 0.0 |
| Total (excl. yr 6)† | | 40 | 161 | 23 | 0 |
| Average (excl. yr 6)† | | 8 | 32.2 | 4.6 | 0 |

†*for direct comparison with 3(a) data*

**Table 4(c)** Table of the number of JJA days when the maximum temperature exceeds 25°C and the minimum exceeds 13°C. (c) Refers to the BRB WISER data for 2031-2036.

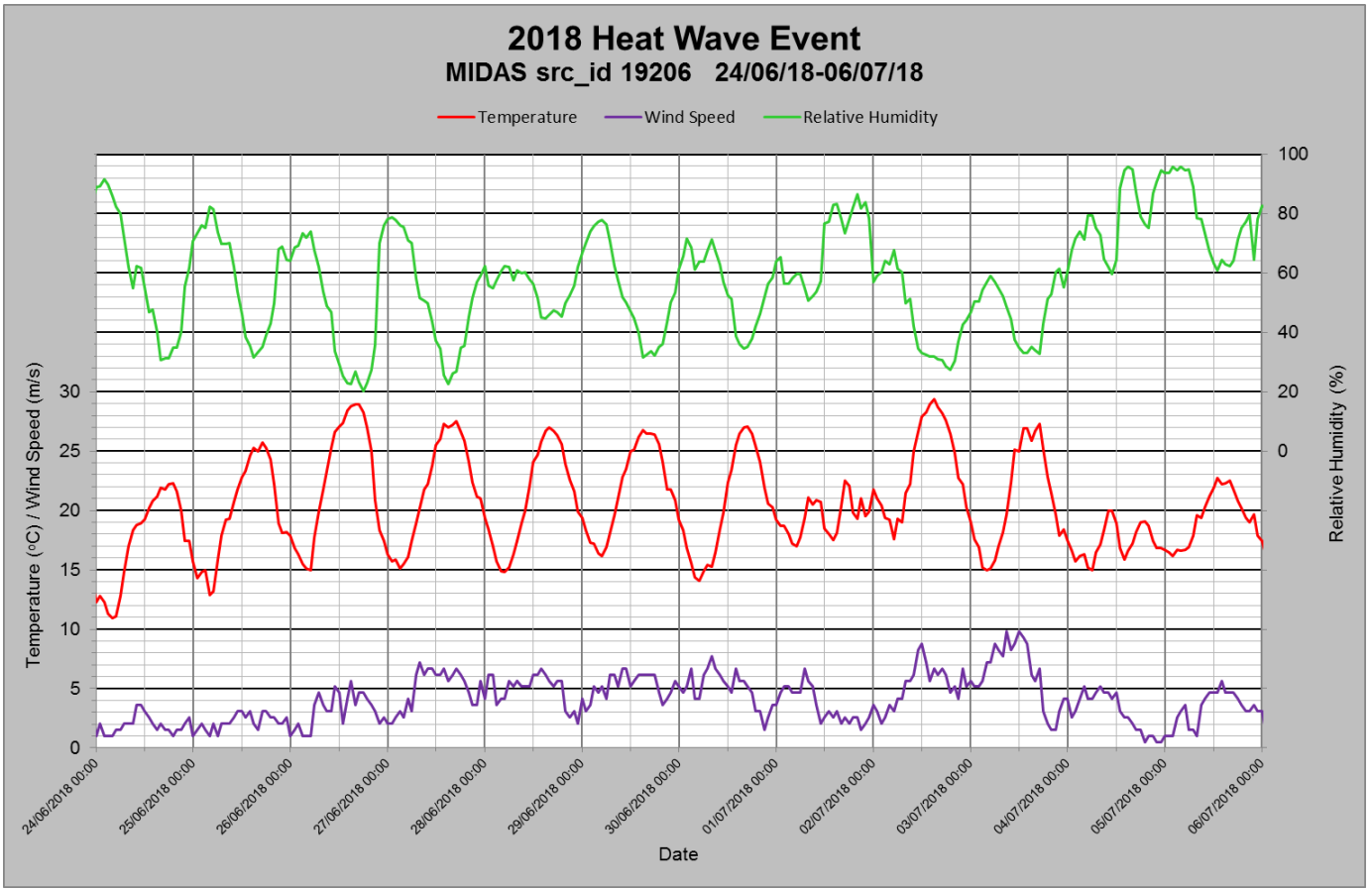

**Figure 4.** 2018 Heat wave event. Hourly temperature, wind speed and humidity data from St Athan weather station (src_id 19206) between 24/06/2018-06/07/2018.

**5. Summary points:**

The WISER, high resolution convective permitting results for 1990-1995 are compared with observations at two UK coastal sites, and compared with the 2031-2036 simulations.

For the control experiment comparison, there is an underestimate / offset of the mean summer (JJA) *maximum* daily temperatures by 1.8°C for BRB, on the east coast of southern UK and 0.5°C for HPC, on the west coast of southern UK. The simulations also underestimate the mean summer *minimum* temperatures by 1.8°C and 2.8°C for BRB and HPC respectively.

For the relative change between the model results between 1990-1955 and 2031-2036, the model suggests a 1.2°C and 1.1°C increase in mean summer *maximum* daily temperatures for BRB and HPC.

Combining these two offsets with the future changes, a simple linear and conservative projection would give an increase of more than 25% of days each year when the summer maximum temperature > 25°C, over 60% increase of events when the minimum temperature is exceeded, combining to produce at least 10 days every year when heat wave type conditions would apply.

For the annual averages, the annual maximum / (minimum) temperature rises are 1.2°C / (0.1°C) and 0.5°C / (0.6°C) for BRB and HPC which are larger values than those predicted in the 2013 IPCC assessment [*Collins et al 2013*], but consistent with the UKCP18 maximum temperatures and increased number of hot days [*Met Office 2019(b)*].

**Acknowledgements:**

The authors acknowledge the use of the ARCHER supercomputer following a UK EPSRC leadership award, NERC computer resources at Jasmin and on the CRAY and resources on the NCAR Yellowstone computer facility. NCAR is sponsored by the National Science Foundation. The data set is available under open access for research purposes. It is currently stored on the National Environment Research Council JASMIN and RDF facilities and can be obtained in the short term by contact the Authors Alan Gadian or Ralph Burton. A data paper is in the process of being submitted for publication, in which the metadata will be described in a doi allocated. The authors also acknowledge the support and funding of the Office for Nuclear Regulation and the National Centre for Atmospheric Sciences.

**Disclaimer:**